\pdfoutput=1
\documentclass[aps,prd,twocolumn,nofootinbib]{revtex4-2}

\usepackage[T1]{fontenc}
\usepackage{lmodern}

\usepackage{amsmath,amssymb}
\usepackage{siunitx}
\usepackage{xcolor,xspace}
\usepackage[colorlinks=true,linkcolor=black,urlcolor=blue,citecolor=red]{hyperref}
\usepackage{graphicx}
\usepackage{hyperref}
\usepackage[capitalise]{cleveref}
\usepackage{booktabs}
\usepackage{multirow,bigdelim}

\newcommand{\updated}[1]{#1}

\newcommand{\order}[1]{\mathcal{O}(#1)}
\newcommand{\lagr}{\mathcal{L}}

\newcommand{\alphaEM}{\alpha_\text{em}}

\newcommand{\newQ}{\mathcal{Q}}
\newcommand{\NQ}{N_\newQ}
\newcommand{\mQ}{m_\newQ}
\newcommand{\gr}[1]{\mathrm{#1}}
\newcommand{\dynk}{T}
\newcommand{\UPQ}{\gr{U}(1)_\text{PQ}}
\newcommand{\NDW}{N_\text{\tiny DW}}

\newcommand{\axion}{a}
\newcommand{\fa}{f_\axion}
\newcommand{\Cagamma}{C_{\axion\gamma\gamma}}
\newcommand{\gagamma}{g_{\axion\gamma\gamma}}

\newcommand{\dd}{\mathrm{d}}
\newcommand{\tr}{\mathrm{tr}}
\newcommand{\ee}{\mathrm{e}}

\newcommand{\pref}{\emph{preferred}\xspace}
\newcommand{\Ntot}{\ensuremath{N_\text{tot}}\xspace}

\begin{document}

\title{Anomaly Ratio Distributions of \updated{Hadronic} Axion Models with Multiple Heavy Quarks}

\author{Vaisakh Plakkot}
\email{vaisakh.plakkot@stud.uni-goettingen.de}
\author{Sebastian Hoof}
\email{hoof@uni-goettingen.de}
\affiliation{Institut f\"{u}r Astrophysik, Georg-August-Universit\"{a}t {G\"{o}ttingen}, Friedrich-Hund-Platz~1, 37077\ {G\"{o}ttingen}, Germany}
\date{September~2021}

\begin{abstract}
We consider \updated{hadronic} axion models that extend the Standard Model by one complex scalar field and one or more new heavy quarks, i.e.\ $\NQ \geq 1$. We review previously suggested selection criteria as well as categorize and catalog all possible models for $\NQ \leq 9$. In particular, allowing for $\NQ > 1$ can introduce models that spoil the axion solution of the strong CP problem. Demanding that Landau poles do not appear below some energy scale limits the number of preferred models to a finite number. For our choice of criteria, we find that $\NQ \leq 28$ and only 820~different anomaly ratios~$E/N$ exist~(443 when considering additive representations, 12 when all new quarks transform under the same representation). We analyze the ensuing $E/N$ distributions, which can be used to construct informative priors on the axion-photon coupling. The \updated{hadronic} axion model band may be defined as the central region of one of these distributions, and we show how the band for equally probable, preferred models compares to present and future experimental constraints.
\end{abstract}

\maketitle

\section{Introduction}\label{sec:intro}
QCD axions~\cite{1978_weinberg_axion,1978_wilczek_axion}, initially proposed as a solution to the strong~CP problem~\cite{1977_pq_axion1,1977_pq_axion2}, are excellent cold dark matter~(DM) candidates~\cite{Preskill:1982cy,Abbott:1982af,Dine:1982ah,Turner:1983he,Turner:1985si}. Numerous experimental searches are currently underway to find such particles~\cite{1801.08127}. One major challenge of axion detection is that the axion mass is set by an unknown parameter, the axion decay constant~$\fa$, which can range across many orders of magnitude. Moreover, the axion's interactions with the Standard Model~(SM) are usually model-dependent, and a UV~axion model has to be constructed in order to determine the exact relationship of $\fa$ and the axion couplings.

One class of such UV models are \updated{hadronic~(also called KSVZ-type)} axion models~\cite{1979_kim_ksvz,1980_shifman_ksvz}, which extend the SM by a new complex scalar field and $\NQ \geq 1$ heavy, exotic quarks. For a given value of $\NQ$ there exist multiple, discrete models, which trace out lines in the axion mass and axion-photon coupling parameter space. The locations of these lines are determined by the anomaly ratio $E/N$ and a model-independent contribution from axion-meson mixing.

To map and restrict the resulting landscape of axion models, it has been suggested that phenomenological selection criteria can be used to single out \pref models~\cite{1610.07593,1705.05370}. This allows us to restrict the parameter space and helps experiments to assess their sensitivity requirements. However, so far only the case of $\NQ = 1$ has been fully cataloged, which is why we want to study models with $\NQ > 1$ as far as this is feasible. First, we summarize the construction of KSVZ-type axion models and phenomenological selection criteria in Secs.~\ref{sec:ksvz_models} and~\ref{sec:criteria}. Subsequently, a catalog of all possible models with $\NQ \leq 9$ is presented and the resulting $E/N$~distributions are discussed. We catalog all \pref models, for which we find that the maximum possible number of $\newQ$s is $\NQ = 28$. In Sec.~\ref{sec:scan} we outline how the catalog of models can be used to construct informative prior distributions on~$E/N$. These can be used to define the KSVZ axion model band and we show how it compares to current and future experimental constraints. Finally, we summarize our work and end with some closing remarks. Model catalogs and further supplementary material are available on Zenodo~\cite{Zenodo_KSVZCatalogue}.

\section{Hadronic axion models}\label{sec:ksvz_models}
Let us denote a representation of a particle as $(\mathcal{C},\mathcal{I},\mathcal{Y})$, where $\mathcal{C}$ and $\mathcal{I}$ are the $\gr{SU}(3)_\mathcal{C}$ color and $\gr{SU}(2)_\mathcal{I}$ isospin representations, respectively, while $\mathcal{Y}$ denotes the particle's $\gr{U}(1)_\mathcal{Y}$ hypercharge.

For example, the traditional KSVZ axion model contains a heavy chiral quark $\newQ = \newQ_L + \newQ_R \sim (3,1,0)$, charged under the $\UPQ$ Peccei-Quinn~(PQ) symmetry with charge~$\mathcal{X} = \mathcal{X}_L-\mathcal{X}_R = \pm 1$, and the complex scalar field $\Phi \sim (1,1,0)$ with PQ~charge normalized to $\mathcal{X}_\Phi = 1$. All SM fields are uncharged under the PQ symmetry in the KSVZ model, and the relevant part of the Lagrangian is
\begin{align}
    \lagr \supset \ &i\,\overline{\newQ}\,\gamma^\mu D_\mu \newQ - (y_\newQ \overline{\newQ}_L \newQ_R \Phi + \text{h.c.}) \nonumber \\
    &- \lambda_\Phi \left( |\Phi|^2 - \frac{v_a^2}{2} \right)^2 , \label{eq:lagr} 
\end{align}
where $y_\newQ$ is the Yukawa coupling constant and the last term is a potential for the complex scalar field with order parameter~$v_a$. The Lagrangian is invariant under a chiral $\UPQ$ transformation $\Phi \mapsto \ee^{i \alpha}\Phi$, \updated{$\newQ_{L/R} \mapsto \ee^{\pm i \alpha/2}\newQ_{L/R}$}. The field $\Phi$ attains a non-zero value at the minimum of the potential, resulting in a spontaneously broken PQ symmetry. Expanding $\Phi$ around its vacuum expectation value gives the axion as the corresponding angular degree of freedom, with value in the interval $[0, 2 \pi v_a)$. The mass of $\newQ$ is then $m_\newQ = y_\newQ v_a/\sqrt{2}$.

Performing a chiral \updated{$\gr{U}(1)$ transformation such that \mbox{$\newQ_{L/R} \mapsto \ee^{\pm ia/(2v_a)}\newQ_{L/R}$}}, the mass term for $\newQ$ can be made independent of the axion field phase. This transformation adds an anomalous $G\widetilde{G}$ term to Eq.~\eqref{eq:lagr} as well as an $F\widetilde{F}$ term, where $G$ and $F$ are the gluon and photon field strength tensors, respectively, and the tilde denotes their duals. With the electromagnetic~(EM) and color anomaly contributions due to the $\UPQ$ charged quarks labeled $E$ and $N$ respectively, the coupling terms become
\begin{align}
    \lagr &\supset \frac{N \alpha_\text{s}}{4 \pi}\frac{a}{v_a}G \widetilde{G} + \frac{E \alphaEM}{4 \pi}\frac{a}{v_a}F\widetilde{F} \nonumber \\
    &= \frac{\alpha_\text{s}}{8\pi \fa}a G \widetilde{G} + \frac{\alphaEM}{8 \pi \fa}\frac{E}{N}a F\widetilde{F}\, , 
    \label{eq:couplingLag}
\end{align}
where $\fa = v_a/(2N)$. The axion-photon coupling is thus parameterized by the anomaly ratio $E/N$ alone.

More precisely, the mass and coupling to photons for QCD~axion models are given by~\cite{1511.02867,1812.01008}
\begin{align}
    m_a &= \frac{\chi_0^2}{\fa} = \SI{5.69(5)}{\micro\eV} \left(\frac{\SI{e12}{\GeV}}{\fa}\right) \, , \label{eq:axion_mass} \\
    \gagamma &=  \frac{\alphaEM}{2\pi\fa} \, \Cagamma = \frac{\alphaEM}{2\pi\fa} \left[\frac{E}{N} - C_{a\gamma\gamma}^{(0)}\right] \nonumber \\
    &= \frac{\alphaEM}{2\pi\fa} \left[\frac{E}{N} - (\num{1.92(4)})\right] \, . \label{eq:axion_gagg}
\end{align}

For some representation~$r$ under which the heavy quark~$\newQ$ in the KSVZ axion model transforms, the EM and color anomalies can be calculated as
\begin{subequations}
\begin{align}
    E &= \mathcal{X} \, d(\mathcal{C}) \,\tr(q^2) \nonumber \\
      &= \mathcal{X} \, d(\mathcal{C}) \, d(\mathcal{I}) \left(\frac{d(\mathcal{I})^2-1}{12}+\mathcal{Y}^2\right) \, , \label{eq:E}\\
    N &= \mathcal{X} \, d(\mathcal{I}) \, \dynk(\mathcal{C}) \, ,
    \label{eq:N}
\end{align}
\end{subequations}
where $d(\cdot)$ denotes the dimension of a representation, $q = \mathcal{I}^{(3)} - \mathcal{Y}$ is the EM charge of $\newQ$, and $\dynk(\mathcal{C})$ is the $\gr{SU}(3)_\mathcal{C}$ Dynkin index~(see Ref.~\cite{1981_Slansky_Review}).

In KSVZ-type models, only~$\newQ$ is charged under the PQ symmetry (apart from $\Phi$) and e.g.\ for $\newQ \sim (3,1,0)$ we have $N = \mathcal{X}/2$ and $E = 3\mathcal{X}\,\tr(q^2)$, using that $T(3) = 1/2$. In general, one finds for a single~$\newQ$ that
\begin{equation}
    \frac{E}{N} = 6 \, \tr (q^2) = 6 q^2 \, ,
\end{equation}
where the last equality holds only when $\newQ$ is a singlet under $\gr{SU}(2)_\mathcal{I}$. This e.g.\ leads to the well-known result that the original KSVZ model has $E/N = 0$.

When considering models with multiple~$\newQ_i$, which have representations $r_i$ and anomaly coefficients $E_i$ and $N_i$ given by Eqs.~\eqref{eq:E} and~\eqref{eq:N}, respectively, the overall anomaly ratio is simply
\begin{equation}
    \frac{E}{N} = \frac{\sum_i E_i}{\sum_i N_i} \, ,
\end{equation}
where the index~$i$ runs over the different quarks, labeled $i=1,\dots,n$.

Note that, when labeling a tuple of $\newQ$s in a model, there exists a ``relabeling symmetry.'' For example, assume that two $\newQ$s  with the same $\UPQ$ charge respectively transform under representations~$r_1$ and~$r_2$, denoted by $r_1 \oplus r_2$. Then there is an equivalency relation such that $r_1 \oplus r_2 \sim r_2 \oplus r_1$, in the sense that they trivially give the same anomaly ratio~$E/N$. Similarly, we can also consider combinations of representations with ``$\ominus$'', the symbol we use to denote $\newQ$s with opposite $\UPQ$ charges such that $r_i \ominus r_j \Rightarrow \mathcal{X}_i = -\mathcal{X}_j$. Here we have e.g.\ $r_1 \oplus r_2 \ominus r_2 \sim r_1 \ominus r_2 \oplus r_2 \sim r_2 \ominus \left( r_1 \oplus r_2 \right)$, as all three models trivially give the same overall anomaly ratio.

The relabeling symmetry allows us to simplify the presentation of the catalog, and we refer to a list of models where this symmetry has been accounted for as ``non-equivalent.'' It may also play a role in the statistical interpretation of the catalog: if not all $\newQ$s are indistinguishable, the multiplicity arising from the equivalency relation must be taken into account. We comment on this further in \cref{sec:priors}.

\section{Phenomenological selection criteria}\label{sec:criteria}
Let us now review the various selection criteria for \pref axion models, most of which have already been proposed and discussed extensively in Refs.~\cite{1610.07593,1705.05370}. Here, we focus on the applicability in the pre-\ and post-inflationary PQ~symmetry breaking scenarios and observe that $\NQ~>~1$ allows for the existence of a new criterion related to the axion's ability to solve the strong~CP problem.

\subsection{Dark matter constraints}
A natural requirement is to demand that axions do not produce more DM than the observed amount, $\Omega_\text{c}h^2 \lesssim 0.12$~\cite{1807.06209}. For QCD~axions this results in an upper bound on~$\fa$ and previous studies of \pref axion models used $\fa < \SI{5e11}{\GeV}$~\cite{1610.07593,1705.05370}, assuming a post-inflationary cosmology with realignment axion production. Let us extend this discussion and make a few comments regarding the different cosmological scenarios and their impact on the $\fa$~bound.

First, in the pre-inflationary PQ~symmetry breaking scenario, the initial misalignment angle of the axion field, denoted by $\theta_\text{i}$, is a random variable. Since any topological defects are inflated away, realignment production is the only relevant contribution and the limit on~$\fa$ depends on its ``naturalness.'' While this is not a uniquely defined concept, using the usual assumption of uniformly distributed angles, $\theta_\text{i} \sim \mathcal{U(-\pi,\pi)}$ the code developed in Ref.~\cite{1810.07192} finds $\fa < \SI{4e12}{\GeV}$ for the 95\% credible region of posterior density.\footnote{Note that we used a prior of $\log_{10}(\fa/\si{GeV}) \sim \mathcal{U}(6,16)$, which introduces some prior dependence, and also included QCD nuisance parameters~\cite{1810.07192}.} This limit on~$\fa$ effectively relies on the naturalness being encoded automatically in prior on~$\theta_\text{i}$.

Second, when topological defects can be neglected in the post-inflationary symmetry breaking, the relic axion density is determined by an average of misalignment angles over many causally-disconnected patches. This corresponds to the benchmark scenario of Refs.~\cite{1610.07593,1705.05370}. Again using the code developed in Ref.~\cite{1810.07192}, we obtain $\fa < \SI{2e11}{\GeV}$~(at the 95\%~CL).

The third and last case is the post-inflationary scenario including a significant contribution from topological defects i.e.\ cosmic strings and domain walls~(DWs). In fact, recent studies indicate that the production of axions via topological defects dominates the vacuum realignment production~\cite{1806.04677,2007.04990}. For models with domain wall number $\NDW \equiv 2N = 1$~(cf.\ \cref{sec:criterionE}), the authors find that $\fa \lesssim \SI{e10}{\GeV}$, while models with $\NDW > 1$ reduce the value of~$\fa$ by a factor~$\mathcal{O}(\NDW)$~\cite{2007.04990}. For the \pref models considered in this work, $\NDW \leq 28$ such that the bound might be loosened to about $\fa \lesssim \SI{3e8}{\GeV}$. It should be noted that these results rely on extrapolating the outcome of numerical simulations more than 60~orders of magnitude, and they hence are potentially subject to large systematic uncertainties.

In summary, the upper limit on~$\fa$, and hence the results presented in what follows, very much depend on the cosmological scenario at hand. To simplify the discussion, to avoid the potentially large uncertainties mentioned above, and to better compare with previous work of Ref.~\cite{1705.05370}, we also adopt $\fa < \SI{5e11}{\GeV}$.

However, we stress again that a different choice of~$\fa$ will affect the number of \pref models, as $\fa$ is one of the factors that determines the value of~$\mQ$. This is because $m_\newQ = y_\newQ \, v_a/\sqrt{2} = y_\newQ \, \NDW \fa / \sqrt{2}$, such that $\fa$ provides an upper bound on $m_\newQ$. Moreover, a universal bound on the $m_\newQ$~(up to the Yukawa couplings) requires that all $\newQ$s are coupled to the $\Phi$~field in the same way to get a single $v_a$ parameter. So long as the coupling $y_\newQ \sim \mathcal{O}(1)$ or lower, the upper bound on \updated{$\fa = v_a/\NDW$} is indeed an upper limit to $m_\newQ$. Larger values of the coupling require fine-tuning of parameters, and are hence deemed undesirable from a theoretical viewpoint. In what follows, we choose $\mQ = \SI{5e11}{\GeV}$ as a conservative value for all $\newQ$ masses~(see Sec.~\ref{sec:lp} for more details on the influence on Landau pole constraints).

Finally, note that the $\newQ$s themselves contribute to the matter content in the Universe, and we need to consider the possibility that their abundance exceeds $\Omega_\text{c}h^2$. Since this issue can be avoided if the lifetime of the quarks is short enough, we discuss this in the next section.

\subsection{Lifetimes}\label{sec:lifetimes}
Other than the possibility that the $\newQ$s' abundances exceed $\Omega_\text{c}h^2$, there also exist additional experimental and observational constraints, which have already been discussed before~\cite{1610.07593,1705.05370}.

To avoid the DM constraints, we require the $\newQ$s to decay into SM particles with a reasonably low lifetime. Heavy quarks with $m_\newQ \gg \SI{1}{\TeV}$ and lifetimes $\SI{0.01}{\s} < \tau_\newQ < \SI{e12}{\s}$ are severely constrained, as they would also affect Big Bang Nucleosynthesis and observations of the Cosmic Microwave Background~\cite{BBNcon,CMBcon}. Fermi-LAT excludes $\SI{e13}{\s} < \tau_\newQ < \SI{e26}{\s}$, thus excluding lifetimes greater than even the age of the Universe ($\sim \SI{e17}{\s}$)~\cite{Fermilat}. As a result, for heavy quarks ($m_\newQ \gg \SI{1}{\TeV}$), only representations with $\tau_\newQ < \SI{e-2}{\s}$ are considered to be a part of the \pref window. Lighter relics would be excluded from experimental bounds e.g.\ at the LHC~\cite{Jager:2018ecz}.

Such a constraint on the $\newQ$ lifetime, when applied to the heavy quark decay rate, translates to restrictions on the dimensionality of the possible $\newQ$ to SM fermion decay operators. With $m_\newQ \lesssim \SI{5e11}{\GeV}$, the lifetime constraints in turn constrain operators to have dimensions $d\leq~5$~\cite{1610.07593,1705.05370}.
This implies a total of 20 possible representations for $\newQ$, all charged under $\gr{SU}(3)_\mathcal{C}$ and $\gr{U}(1)_\mathcal{Y}$.
The lifetime constraint has no further consequence on cases with $\NQ > 1$ under the assumption that the different $\newQ_i$ do not interact among themselves or decay into particles other than SM fermions.

As noted before~\cite{1705.05370}, the lifetime constraints are \updated{typically} not required in the pre-inflationary PQ~symmetry breaking scenario. \updated{This is because the $\newQ$s can get diluted by inflation, which prevents them from becoming cosmologically dangerous relics after they freeze out. Without these constraints,} many more models with even higher-dimensional operators \updated{can} exist, and restricting ourselves to at most five-dimensional operators therefore only becomes an assumption in this case.

\subsection{Failure to solve the strong CP problem}\label{sec:CPfailure}
This criterion is specific to models with $\NQ > 1$ that allow the $\newQ$s to have opposite $\UPQ$ charges.
It is clear from \cref{eq:N} that the addition of multiple heavy quarks can lead to a smaller overall $N$ than the individual $N_i$, but only when one or more of the quarks have a (relative) negative $\UPQ$ charge.
In some cases a total cancellation of the $N_i$ terms occurs~($N = 0$).
While these models give rise to massless axion-like particles with a coupling to photons governed by~$E$, they do not solve the strong~CP problem: as can be seen from \cref{eq:couplingLag}, $N = 0$ means that there is no $G\widetilde{G}$ contribution in the Lagrangian. Considering that the primary objective of QCD axion models is to solve the strong CP problem, we propose that only models with $N \neq 0$ should be considered \pref.

\subsection{Landau poles}\label{sec:lp}
The single most powerful criterion amongst the ones proposed by Refs.~\cite{1610.07593,1705.05370} in the context of this work comes from the observation that representations with large $\mathcal{C}$, $\mathcal{I}$, or $\mathcal{Y}$ can induce Landau poles~(LPs) at energies well below the Planck mass. At an LP, the value of a coupling mathematically tends to infinity, signaling a breakdown of the theory. Since quantum gravity effects are only expected to appear at energies near the Planck mass, a breakdown of the theory before that point can be regarded as problematic or undesirable.

It has thus been proposed that \pref models have LPs at energy scales $\Lambda_\text{LP} \gtrsim \SI{e18}{\GeV}$. From the 20~representations mentioned previously, only 15~fulfil this criterion~\cite{1610.07593,1705.05370}; we refer to these as ``LP-allowed'' models and label them $r_1$ to $r_{15}$ (as per Table~II in Ref.~\cite{1610.07593}).

The running of the couplings are computed at two-loop level with the renormalization group equation~\cite{Machacek:1983tz,DiLuzio:2015oha}
\begin{align}
    \frac{\dd}{\dd\mathrm{t}}\alpha_i^{-1} &= -a_i - \frac{b_{ij}}{4\pi}\alpha_j \, \label{eq:rge} \, ,
\end{align}
where
\begin{widetext}
\begin{subequations}
\begin{align}
    a_i &= -\frac{11}{3} \, C_2(G_i) + \frac{4}{3}\sum_F \kappa \, \dynk(F_i) + \frac{1}{3} \sum_S \eta \, \dynk(S_i) \, ,\\
    b_{ij} &= \left[-\frac{34}{3} \, \big(C_2(G_i)\big)^2 + \sum_F \left( 4 C_2(F_i) + \frac{20}{3}C_2(G_i) \right)\kappa \, \dynk(F_i) + \sum_S \left(4C_2(S_i)+\frac{2}{3}C_2(G_i)\right)\eta \, \dynk(S_i) \right] \delta_{ij} \nonumber \\ &+ 4 \left(1-\delta_{ij} \right) \left[ \sum_F \kappa \, C_2(F_j) \, \dynk(F_i) + \sum_S \eta \, C_2(S_j) \, \dynk(S_i) \right] \, , \label{eq:betafns}
\end{align}
\end{subequations}
\end{widetext}
with $i,j\in \{1,2,3\}$ for the three gauge groups, $\alpha_i = g_i^2/4\pi$, $\mathrm{t} = \frac{1}{2\pi}\ln(\mu/m_Z)$ for energy scale $\mu$ and $Z$~boson mass $m_Z$, while $a_i$ and $b_i$ are the one-\ and two-loop beta functions. $C_2$~and $\dynk$ are the quadratic Casimir and Dynkin indices of the corresponding gauge group, respectively, and $F$ and $S$ denote fermionic and scalar fields. $G_i$ denotes the adjoint representation of the gauge group, and $\kappa=\frac{1}{2},1$ for Weyl and Dirac fermions, while $\eta=1$ for complex scalars.\footnote{The case of $\eta=\frac{1}{2}$ for real scalars is not relevant for the present study. Also note that the expression for $b_{ij}$ in Ref.~\cite{DiLuzio:2015oha} is slightly erroneous since the second term applies only to the non-diagonal elements of $b_{ij}$, as found when comparing with the SM beta functions in Ref.~\cite{Machacek:1983tz}.} Adding multiple $\newQ$s to the theory increases the coefficients of beta functions through the fermionic terms. As a consequence, the couplings diverge faster i.e.\ induce LPs at lower energy scales, as has been anticipated before~\cite{1705.05370}.

Since the addition of more particles with a given representation into a gauge theory only worsens the running of the corresponding gauge coupling, it is possible to find the number of copies of a particle that can be included in the theory before it induces an LP below $\SI{e18}{\GeV}$.
This drastically reduces the number of LP-allowed combinations possible.
Integrating all $\newQ_i$ in at $m_\newQ = \SI{5e11}{\GeV}$, we find that there are 59,066 non-equivalent combinations of $\newQ_i$ from the representations  $r_1, r_2, \dots ,r_{15}$ that do not induce LPs below \SI{e18}{\GeV}. 

As the $\newQ_i$ contribute to the beta functions above the energy scale $m_\newQ$, the running of the gauge coupling begins to deviate from the SM only at this scale. Different values of $m_\newQ$ are bound to produce different results for the LPs; the lower $m_\newQ$ is, the earlier an LP appears. As an example, consider $m_\newQ = \SI{e10}{\GeV}$: for $\NQ = 3$, we find that 888 models are \pref~(they have $\Lambda_\text{LP} > \SI{e18}{\GeV}$ and $N \neq 0$ as per the discussion in~\ref{sec:CPfailure}), compared to 1,442 models when $m_\newQ = \SI{5e11}{\GeV}$. Furthermore, we use the same mass for all $\newQ_i$ in the models, which may not be the case in reality (due to different $y_{\newQ_i}$ or e.g.\ in multi-axion models). However, setting the masses to the highest possible value in the \pref window allows us to keep the number of disfavored models to a minimum. Without further information on the values of $\fa$ and individual $m_{\newQ_i}$, excluding fewer models may be advantageous in the sense of presenting a more inclusive $E/N$ catalog.

\subsection{Other interesting model properties}\label{sec:criterionE}
Let us summarize a few other possible model properties, already discussed in Refs.~\cite{1610.07593,1705.05370}. 

Since the axion is the angular degree of freedom of the PQ scalar field, it has a periodic potential and several degenerate vacua, given by the domain wall number $\NDW = 2N$. During PQ symmetry breaking, the axion field can settle into any of these degenerate minima in different Hubble patches, giving rise to domain walls. The energy density contained in such topological defects can far exceed the energy density of the Universe~\cite{Domainwallproblem} in the post-inflationary PQ breaking scenario. However, in models with $\NDW = 1$, the string-domain wall configuration would be unstable~\cite{Barr:1986hs}, which presents a possible solution and makes $\NDW = 1$ a desirable property of such models.

However, the DW problem can be avoided by allowing for a soft breaking of the PQ symmetry~\cite{Domainwallproblem}. Moreover, in a pre-inflationary PQ symmetry breaking scenario, the patches and the topological defects are inflated away~\cite{Kim:1986ax}. In line with Refs.~\cite{1610.07593,1705.05370}, we therefore do not impose this criterion.

Among the 15~LP-allowed representations, only two have $\NDW = 1$. When all $\newQ_i$ have the same $\UPQ$ charges, such a restriction would forbid any models with multiple heavy quarks. With this in mind, a constraint on $\NDW$ is not used to exclude $\NQ > 1$ models. In cases where the $\newQ_i$ are permitted to have opposite $\UPQ$ charges, more complicated models with $\NDW = 1$ can be built by choosing the $\newQ_i$ such that $\sum_i N_i = 1/2$. Even then, the number of such models is few in comparison to the whole set of LP-allowed models.

Another intriguing property is the unification of the gauge couplings due to the presence of the $\newQ$s. The authors of Refs.~\cite{1610.07593,1705.05370} note that one of the 15 LP-allowed representations induces a significant improvement in unification. While we do not investigate this further, we expect to find more models that improve unification for higher $\NQ$, which might be an interesting topic for a future study.

\section{Model catalog and anomaly ratio distributions}\label{sec:distributions}
\begin{table*}
    \caption{Selected statistics for the complete set of models with $\NQ \leq 9$.
    We include information about the $E/N$ ratios that give rise to the largest axion-photon coupling i.e.\ $\widehat{E/N} \equiv \mathrm{argmax}_{E/N}(|E/N - 1.92|)$, photophobic models~($|E/N - 1.92| < 0.04$), and \pref (LP-allowed and $N \neq 0$) models.
    }
    \centering
    \setlength{\tabcolsep}{10pt}
    \begin{tabular}{crcrrrcr}
    \toprule
    $\NQ$ & \multicolumn{1}{l}{Total \#models} & $\widehat{E/N}$  & \multicolumn{1}{l}{LP-allowed} &
    \multicolumn{1}{r}{$N\neq 0$} &
    \multicolumn{1}{l}{\#\pref} &
    $\widehat{E/N}$ &
    \multicolumn{1}{l}{photophobic} \\
    \cmidrule(lr){4-5}\cmidrule(lr){7-8}
    & & & \multicolumn{2}{c}{fraction of total [\%]} & \multicolumn{1}{c}{} & \multicolumn{2}{c}{among \pref} \\
    \midrule
    1 & 20 & $\phantom{-00}44/3$ & 75.00 &  100.00 &  15 & $\phantom{-0}44/3$ & 0.00\% \\
    2 & 420 & \phantom{0}$-184/3$ & 49.52  & 91.67 & 189 & $\phantom{-}122/3$ &1.59\% \\
    3 & 5,740 & $\phantom{-0}368/3$ & 25.98  & 97.40 & 1,442 & $\phantom{-}170/3$ &1.11\% \\
    4 & 61,810 & \phantom{0}$-538/3$ & 11.60 & 97.37 & 6,905 & $-136/3$ &1.29\% \\
    5 & 543,004 & $\phantom{-0}698/3$ & 4.42 & 98.13  & 23,198 & $-148/3$ &1.27\% \\
    6 & 4,073,300 & \phantom{0}$-928/3$ & 1.50 & 98.32  & 58,958 & $-160/3$ &1.28\% \\
    7 & 26,762,340 & $-1108/3$ & 0.47 & 98.55  & 120,240 & $\phantom{-}164/3$ &1.33\% \\
    8 & 157,233,175 & $\phantom{-}1292/3$ & 0.14 & 98.68  & 207,910 & $-166/3$ &1.34\% \\
    9 & 838,553,320 & $-1312/3$ & 0.04 & 98.79 & 312,360 & $-142/3$ &1.37\% \\
    \bottomrule
    \end{tabular}
    \label{tab:findings}
\end{table*}
\begin{figure}
    \centering
    \includegraphics[width=3.375in]{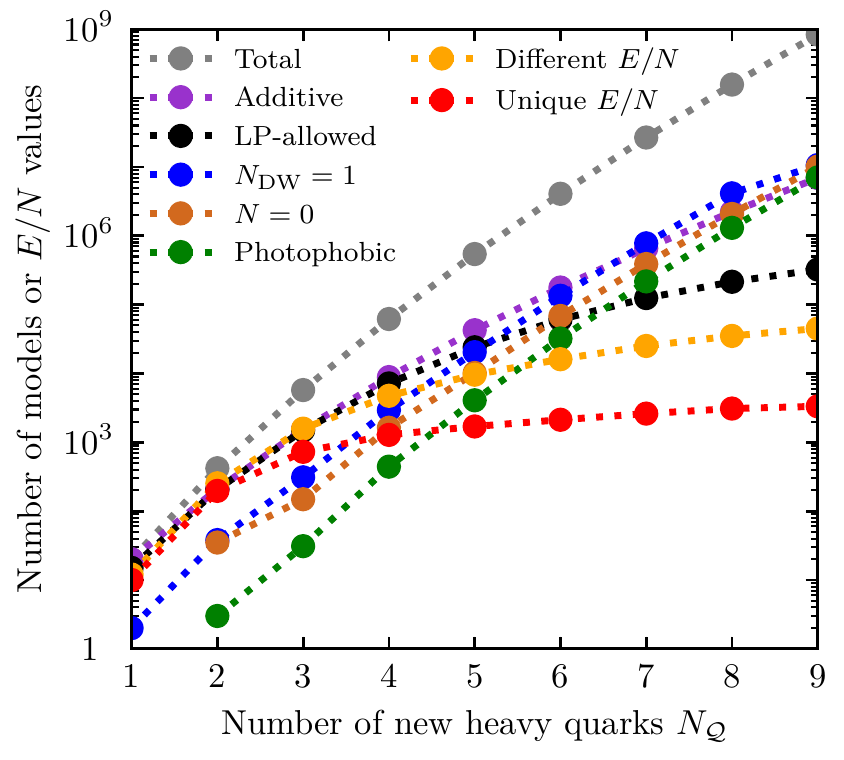}
    \caption{Number of non-equivalent models with different properties as a function of $\NQ$. We show the number of all possible, additive, LP-allowed, $\NDW = 1$, $N = 0$, photophobic ($|E/N - 1.92| < 0.04$) models, as well as the number of different and unique~(no other non-equivalent model has the same $E/N$ value such that the underlying model is uniquely identifiable) $E/N$ values.\label{fig:number_of_models}}
\end{figure}
Let us discuss a few key findings and properties of the model catalog created in this work, which we summarize in \cref{tab:findings} and \cref{fig:number_of_models}.

To structure the discussion, we single out two subsets of the total model space: one where all $\newQ_i$ transform under the same representation and one where the representations are arbitrary but the $\UPQ$ charges of the quarks have the same sign~(we call these ``additive models'').

\subsection{Subset~I. Identical representations}
First, consider the case where only representations of the form $\bigoplus_{j=1}^{\NQ} r_i$ with fixed~$i \in [1,20]$ are allowed. The number of possible models for a given $\NQ$ is then simply $N_r = 20$, such that the total number of models up to and including some $\NQ$ is $\Ntot = N_r \, \NQ$.

Given that all quarks in such models have the same representation and $\UPQ$ charge, only twelve discrete values of $E/N$ are allowed when the LP criterion is taken into account~\cite{1610.07593}. However, the relative distribution is determined by the effect of each representation on the gauge group beta functions. We find that $\bigoplus_{j=1}^{28} r_1$ is the only LP-allowed model for $\NQ = 28$ and that there are in total 79 \pref models in this subset.

\subsection{Subset~II. Allowing different additive representations}
Next, consider the case where we can have arbitrary additive representations, written in such a way that they respect the relabeling symmetry: $\bigoplus_{i=1}^{20} \bigoplus_j^{n_i} r_i$, where $\sum_i n_i = \NQ$ with $n_i \geq 0$. The number of models in this subset \updated{is}

\begin{align}
    N(\NQ) &= \binom{\NQ+N_r-1}{\NQ} \, , \\
    \Ntot &= \sum_n \binom{n+N_r-1}{n} =  \binom{\NQ+N_r}{\NQ} \, .
\end{align}

We find that, after applying the selection criteria, there are 59,066 \pref models for $\NQ \leq 28$. In particular, for $\NQ = 28$, there are only nine LP-allowed models, none of which can be extended by another quark while preserving the criterion. The highest freedom in this subset is found for $\NQ = 10$, where 5,481 models fall in the \pref region.

Among these models, the smallest and largest anomaly ratios are 1/6 and 44/3 respectively, both of which come from $\NQ = 1$ models. The median of the distribution of this set of models is $\mathrm{med}(E/N) \approx 1.87$, indicating that $|\Cagamma| \sim 0$ is a real possibility for a larger fraction of the model space. Indeed, there are several models that have an $E/N$ ratio close to the nominal value of the model-independent parameter $\Cagamma^{(0)}$. We define models as ``photophobic'' if their $E/N$~ratio is within one standard deviation of the nominal $\Cagamma^{(0)}$ value:
\begin{equation}
    \left|E/N - 1.92 \right| < 0.04 \, . \label{eq:photophobic_definition}
\end{equation}

We find that 3,255 models ($\approx 5.5\%$) among the 59,066 non-equivalent models are photophobic. Considering all \pref additive models up to $\NQ \leq 28$, there are 443 different $E/N$ values. Out of these, 28 are unique in the sense that they are uniquely identifiable since their anomaly ratio~$E/N$ is different from any other non-equivalent model.

\subsection{Complete set}
\begin{figure}
    \centering
    \includegraphics[width=3.375in]{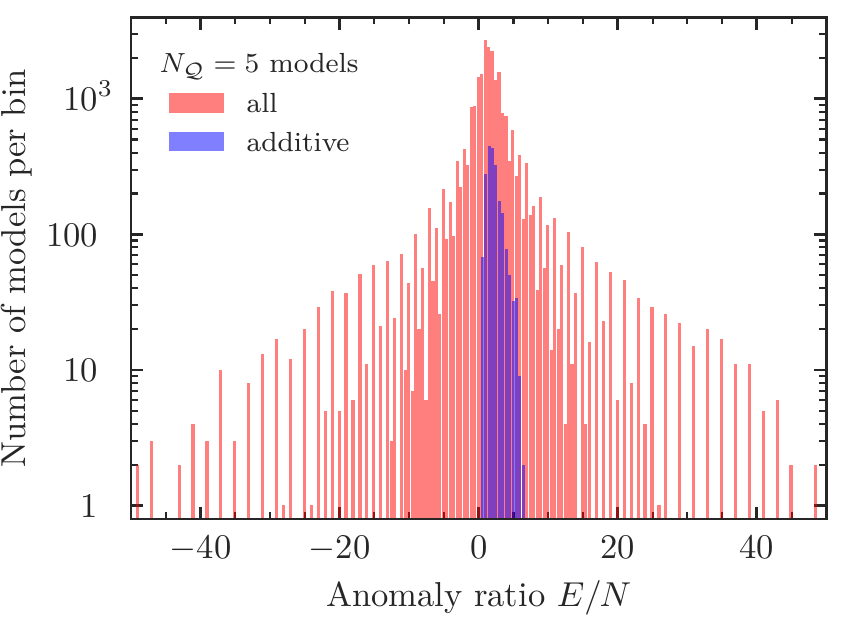}
    \caption{Example histogram of the anomaly ratio $E/N$ for non-equivalent $\NQ = 5$ models. Blue bars correspond to the ``additive'' subset and red bars to the complete set of models i.e.\ also allowing for opposite $\UPQ$ charges.
    }
\label{fig:histogram}
\end{figure}
Finally, let us comment on the complete set of possible models where we may also subtract representations, denoted by~``$\ominus$.'' Allowing $\UPQ$ charges to have one of the two possible values for each $\newQ_i$, we open the window to a much wider range of possible $E/N$ values. In particular, the anomaly ratio, and thus the axion-photon coupling, can become negative~(see \cref{fig:histogram}) and, as mentioned before, the solution to the strong~CP problem can be spoilt in models with $N = 0$.

For $n_\oplus + n_\ominus = \NQ$, where $n_\oplus$ and $n_\ominus$ are the number of $\newQ$s with ``positive'' and ``negative'' $\UPQ$ charges,\footnote{We remind the reader that ``positive'' and ``negative'' are only relative concepts, in the sense that we consider two models equivalent if the only difference between them is that the $\UPQ$ charges of \emph{all} quarks get flipped going from one to the other.} respectively, the number of models with $n_\oplus > n_\ominus$ is simply
\begin{align}
    N(n_\oplus, n_\ominus) = \binom{n_\oplus+N_r-1}{n_\oplus} \, \binom{n_\ominus+N_r-1}{n_\ominus} \, .
\end{align}

In the case where $n \equiv n_\oplus = n_\ominus$, accounting for the fact that the anomaly ratio depends on the relative $\UPQ$ charges of the $\newQ$s such that we have an equivalence of the type $(r_i~\oplus~r_j)~\ominus~(r_k~\oplus~r_l)~\sim~(r_k~\oplus~r_l)~\ominus~(r_i~\oplus~r_j)$, we also need to take care not to double-count models exhibiting this symmetry, giving
\begin{align}
    N(n, n) = \frac{1}{2}\,\binom{n+N_r-1}{n}\left[\binom{n+N_r-1}{n}+1\right]\,.
\end{align}
With this, we find that the number of models grows very fast as $\NQ$ increases. This also makes it computationally difficult to compute and store all of the different combinations -- let alone check the criteria for \pref models. We therefore restrict the complete analysis in this case to $\NQ \leq 9$.

The anomaly ratio distribution in the complete set exhibits a peak near zero, and we expect the trend to continue even for larger $\NQ$. However, in general care should be taken when interpreting the ``trends'' visible in \cref{fig:number_of_models}. For example, the number of LP-allowed models will eventually go down again as we move towards $\NQ = 28$, despite the quickly growing total number of possible models. One may speculate that the number of uniquely identifiable $E/N$ ratios could exhibit a similar behavior as the number of LP-allowed models, while the number of different $E/N$ might eventually saturate. 

Allowing for opposite $\UPQ$ charges gives rise to models with large axion-photon coupling; the largest and smallest values of $E/N$ found, $170/3$ and $-166/3$ respectively, give larger $|\Cagamma|$ than what is possible in the previously discussed subsets. Note that the $\NQ = 8$ model for $E/N = -166/3$ ($r_2 \oplus r_2 \oplus r_5 \oplus r_6 \oplus r_7 \ominus r_1 \ominus r_9 \ominus r_9$) was not reported in Refs.~\cite{1610.07593,1705.05370} as giving the highest possible $|\Cagamma|$; instead the authors indicated that $E/N = 170/3$ led to the largest absolute value of the coupling. We find that among the complete set of 5,753,012 \pref models, there are 81,502 photophobic models and 820 different anomaly ratios, with 79 out of those also being from uniquely identifiable models.

\section{Impact on axion searches}\label{sec:scan}
In this section, we discuss possible statistical interpretations of the \updated{hadronic axion} model catalog and show the impact of these on the mass-coupling parameter space.

\subsection{On constructing E/N prior distributions}\label{sec:priors}
The catalog of KSVZ models -- even after applying the selection criteria -- is but a list of \emph{possible} models. It does not inherently contain information about how \emph{probable} each model is. The model with $E/N = -166/3$ gives the largest $|\Cagamma| \approx 57$, which will place an upper bound on the axion-photon coupling and delimit the upper end of the KSVZ axion band. On the other end, complete decoupling with photons~($\Cagamma \approx 0$) is also possible within the theoretical errors. Since any of the models might be realized in Nature, perhaps due to a deeper underlying reason that is not obvious at present, one might be satisfied with this picture.  

However, the boundaries of the band are extreme cases and do not take into account where the bulk of possible models can be found. For example, defining a desired target sensitivity for an experiment becomes non-trivial in the face of $\Cagamma$ potentially being extremely close to zero. We propose instead that covering a certain fraction of all possible models or constructing a prior volume might be more meaningful ways to define such a target.

To directly interpret an $E/N$~histogram as a distribution implicitly makes the assumption that each model is equally likely to be realized in Nature. While this interpretation might be considered ``fair,'' one could argue that models with many $\newQ$s are more ``contrived'' and consequently introduce a weighting factor that penalizes models with $\NQ \gg 1$. This could be achieved with e.g.\ exponential suppression via a weighting factor $\propto \ee^{-\NQ}$, or $\propto 2^{-\NQ}$. Another option could be to choose  models that are minimal extensions~($\NQ = 1$) or similar to the family structure of the SM~($\NQ = 3$ or e.g.\ a weighting $\propto 3^{\NQ}/\NQ!$).

Such consideration are aligned with the Bayesian interpretation of statistics, and will probably meet criticism for this reason. However, as pointed out in Ref.~\cite{1810.07192}, at least in the pre-inflationary PQ~symmetry breaking scenario, which is fundamentally probabilistic in nature, the Bayesian approach is well motivated. Furthermore, Ref.~\cite{1810.07192} also proposed that the discrete nature of KSVZ models should be reflected in the prior choice of $E/N$. Such a physically-motivated prior should further reflect the combinatorics of KSVZ model building by including the multiplicity of $E/N$ ratios. As mentioned at the end of \cref{sec:ksvz_models}, this multiplicity also depends on whether or not the $\newQ_i$ are distinguishable by e.g.\ having different masses.

\begin{figure}
    \centering
    \includegraphics[width=3.375in]{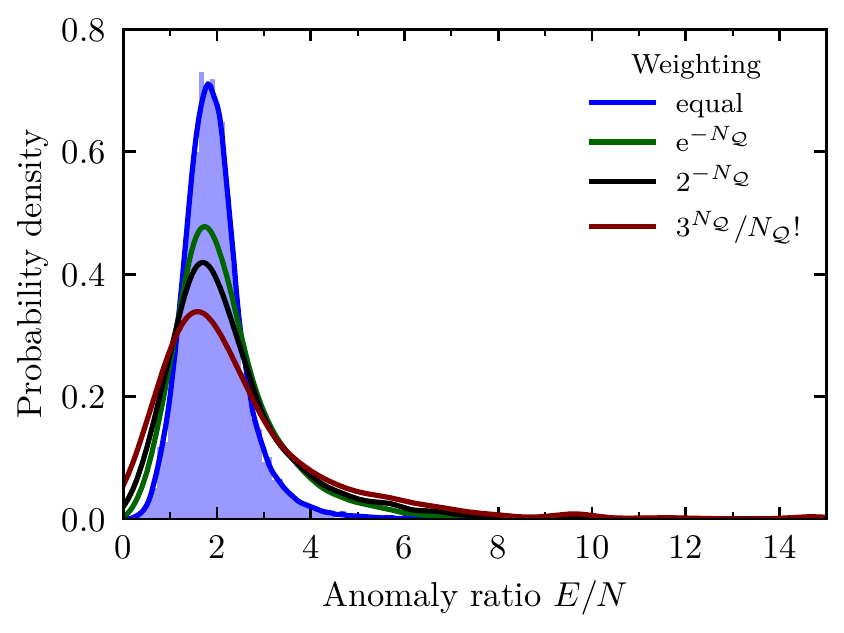}
    \caption{Anomaly ratio distributions for all \pref additive KSVZ models, using different weightings. For equal weighting, we show the underlying histogram~(blue shading) and a smooth Gaussian kernel density estimate of the distribution~(blue line), while for others we only show the latter for simplicity.}
    \label{fig:histogram_fit}
\end{figure}
With this in mind, we show different statistical interpretations of the anomaly ratio in \cref{fig:histogram_fit}. For visualization purposes, we show kernel density estimates of the distributions for different weighting factors mentioned above, while reminding the reader that the underlying histograms and distributions are actually discrete and not continuous.

From \cref{fig:histogram_fit} it becomes clear that the different weightings can change the width of the distribution, introducing a prior dependence in an analysis. However, the modes of the distributions remain around $E/N \sim 2$, which means that a partial cancellation of the axion-photon coupling $\Cagamma$ is typically possible, as already observed in \cref{fig:histogram}.

\subsection{Experimental constraints on \emph{preferred} KSVZ axion models}\label{sec:pheno_study}
\begin{figure*}
    \centering
    \includegraphics[width=5.0625in]{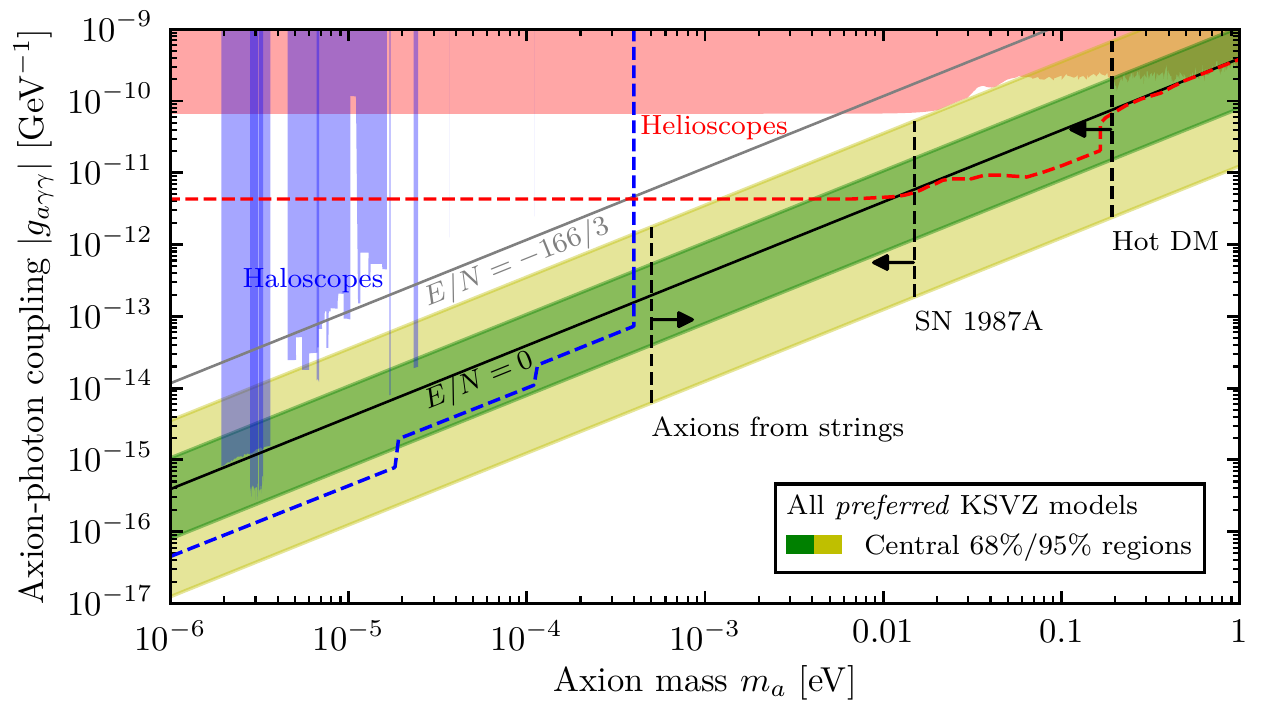}
    \caption{The KSVZ axion band as defined by the 68\% and 95\% central regions of~$|\Cagamma| = |E/N - \Cagamma^{(0)}|$, drawing $E/N$ from a distribution of all \pref KSVZ axion models~(each representation assumed to be equally probable). The grey line marks the highest possible absolute value of the coupling~($E/N = -166/3$), while the black line indicates the classical KSVZ model~($E/N = 0$). For context, we show various present~(shaded regions) and future~(dashed lines) haloscope~(blue) and helioscope~(red) limits and forecasts~\cite{2020_Zenodo_OHare} as well as bounds from hot dark matter~\cite{2011.14704}, energy loss in SN1987A~\cite{1906.11844}, and recent string simulations~\cite{2007.04990}.}
\label{fig:mass_coupling}
\end{figure*}
Of course, a possible partial cancellation of the axion-photon coupling has consequences on the various astrophysical, cosmological, and laboratory searches~(see e.g.\ Ref.~\cite{1801.08127}) for axions. The most powerful analyses combine the results of different experiments to place joint limits on the properties of different types of axions~(e.g.\ Refs.~\cite{1708.02111,1809.06382,1810.07192}).

To investigate this further, consider e.g.\ a prior on $E/N$ where all \pref~(LP-allowed models with $d\leq 5$ operators and $N\neq 0$), non-equivalent KSVZ models are considered equally probable.\footnote{Recall that the $d \leq 5$ condition is due to the lifetime constraints~(see \cref{sec:lifetimes}) in the post-inflationary scenario, while it is only an assumption for the pre-inflationary case~(potentially reasonable for as being a minimal extension of the SM).} We can then generate samples for $\Cagamma = E/N - \Cagamma^{(0)}$, where $E/N$ is drawn from its discrete distribution and $C_{a\gamma\gamma}^{(0)} \sim \mathcal{N}(1.92,\, 0.04)$ i.e.\ follows a normal distribution with mean~1.92 and standard deviation~0.04.

We find that the central 68\% region of the ensuing distribution corresponds to $\updated{|\Cagamma|} \in [0.39,5.22]$, while the 95\% region is $\updated{|\Cagamma|} \in [0.06,17.30]$. The corresponding model bands in the mass-coupling plane are shown in \cref{fig:mass_coupling}, and the bulk of these models can be constrained by present and future experiments. \updated{In fact, while complete cancellation of $\Cagamma$ is possible within the theoretical uncertainty for some $E/N$ values, we find that the bulk of models is at worst somewhat suppressed. This is very encouraging for experimental searches.}

Had we only considered additive models, the 95\% region would be $\updated{|\Cagamma|} \in [0.02,1.67]$, such that the upper end of the band would be lower than the traditional KSVZ model with $E/N = 0$. This can be readily understood from the $E/N$~distributions in \cref{fig:histogram_fit}, whose mode is typically close to $C_{a\gamma\gamma}^{(0)}$ such that the value of $|\gagamma|$ is lower than what would be expected for $|\Cagamma| \sim \order{1}$. In this case, the planned future experiments would \emph{not} be able to probe large parts of the band, indicating that the choice of prior -- even if physically-motivated -- can induce a noticeable impact on the results.

\section{Summary and conclusions}
We provide a catalog of all \updated{hadronic, or KSVZ, axion} models with $\NQ \leq 9$, featuring 1,027,233,129  non-equivalent models in total. When we apply the selection criteria for \pref models, we find a limit of $\NQ \leq 28$ and that only 5,753,012 non-equivalent models with 820 different $E/N$~values exist (59,066 non-equivalent models with 443 different $E/N$~values for additive representations). While relaxing existing or adding new criteria can increase or reduce these numbers, we generically expect that the Landau pole~(LP) criterion will be a powerful tool to limit the number of possible models -- even with modified constraints or in other axion models. This is similar to the Standard Model, where the number of families can also be restricted by demanding that LPs do not appear below some energy scale. We further propose that only models with QCD anomaly $N \neq 0$ be considered \pref.

Our model catalog can be a useful, searchable database for researchers wishing to study the KSVZ axion model space. It allows to e.g.\ make statements about what fraction of possible models a given experiment is sensitive to. We made catalogs, histograms, and example Python scripts available on the Zenodo platform for this purpose~\cite{Zenodo_KSVZCatalogue}.

Some models in the catalog might be considered ``contrived'' as they add many new particles to the theory. Of course, in case of a discovery or if any other appealing reason for a seemingly more complicated models is put forward, this perception might change. In absence of such reasons, the $E/N$~values may be interpreted as statistical distributions, which encode assumptions about the probability of the different models. We generally outlined how prior distributions can be constructed from the catalog and gave concrete examples of such choices.

For the specific choice of equally probable \pref models, we consider the consequences for axion searches and the definition of the KSVZ axion band. Here we suggest that the latter may be defined as the central 95\% region of all models, taking into account uncertainties from the model-independent contribution to the axion-photon coupling. If only ``additive models'' are considered, the bulk of the \pref models can unfortunately not be probed by current or future experiments since the anomaly ratio distributions in this case tend to peak around $E/N \sim 2$. In general, using the discrete $E/N$ distributions improves on unphysical prior choices considered in the past~(e.g.\ Ref.~\cite{1810.07192}).

\updated{Even when ignoring the statistical perspective, it is useful for axion searches to know that the \pref models only admit 820 different $E/N$ values. In case of an axion detection, one may therefore test these discrete models against each other to see which models are most compatible with the detected signal. One could further test them against a generic axion-like particle or other QCD~axion models. In an ideal scenario, this might even allow an experiment to infer the underlying high-energy structure of a model, which highlights the known property of axion models to connect high-energy physics to low-energy observables.}

In summary, the powerful LP~criterion restricts the number of KSVZ models to a finite value. In that sense, the catalog presented here is a complete list of all \pref KSVZ models, which may be used as input for axion searches and forecasts. Since KSVZ models could e.g.\ be extended by also considering multiple complex scalar fields or feature more complex couplings to the SM, and since there are other kinds of QCD~axion models such as the DFSZ-type models, this work presents another step forward in mapping the landscape of all phenomenologically interesting axion models.

\begin{acknowledgments}
We thank \updated{Maximilian Berbig,} Joerg Jaeckel, and David `Doddy' J.~E. Marsh for useful comments and discussions.
This paper is based on results from VP's ongoing M.Sc.\ project.
SH\ is supported by the Alexander von Humboldt Foundation and the German Federal Ministry of Education and Research. We used the Scientific Computing Cluster at GWDG, the joint data center of Max Planck Society for the Advancement of Science~(MPG) and the University of {G\"ottingen}.
\end{acknowledgments}

\bibliographystyle{JHEP_mod}
\bibliography{references}

\end{document}